\documentclass[preprint,showpacs,preprintnumbers,amsmath,amssymb]{revtex4}

\usepackage{graphicx}
\usepackage{dcolumn}
\usepackage{bm}

\begin{document}

\preprint{APS/123-QED}

\title{Reply to ``the Comment on `Modified Scalar-Tensor-Vector Gravity Theory and the Constraint on its Parameters' "}

\author{Xue-Mei Deng$^{1}$}
 \email{xue.mei.deng.x@gmail.com}
\author{Yi Xie$^{2}$}
 \email{yixie@nju.edu.cn}
\author{Tian-Yi Huang$^{2,3}$}
 \email{tyhuang@nju.edu.cn}
 \affiliation{%
$^{1}$Purple Mountain Observatory, Chinese Academy of Sciences,
Nanjing 210008, China\\
$^{2}$Department of Astronomy, Nanjing University, Nanjing 210093,
China\\
$^{3}$Shanghai Astronomical Observatory, Chinese Academy of
Sciences, Shanghai 200030, China }%

\date{\today}

\begin{abstract}
J. W. Moffat and V. T. Toth submitted recently a comment \cite{b7}
on our latest paper ``Modified scalar-tensor-vector gravity theory
and the constraint on its parameters" \cite{b8}. We reply to each of
their comments and justify our work and conclusions. Especially,
their general STVG (MOG) theory has to be modified to fit the modern
precision experiments.
\end{abstract}

\maketitle

In the following we will abbreviate the comment paper of J. W.
Moffat and V. T. Toth \cite{b7} as MT and our paper \cite{b8} as
DXH. We will give point-by-point response to their comments, but may
change the order of the comments in their article.

(1) On the violation of equivalence principle.

Under the condition that the charge of the fifth force is
proportional to its gravitational mass, STVG and MSTVG satisfy only
the {\it Weak Equivalence Principle} (WEP) but not the {\it Einstein
Equivalence Principle} (EEP). The reason is that the vector field
$\phi_{\mu}$ appears in the action of matter $\mathcal{S}_{M}$
besides the matter and the metric (see Eq.(8) in DXH), then the
motion of test particles move not only under the geometry of the
curved space-time. In our paper we stressed that STVG and MSTVG
violate EEP but have never said they violate WEP.

 (2) On the so called effective gravitational constant.

 In post-Newtonian expansion of the solution of the field equations,
 an effective gravitational constant takes the place of the
 Newtonian constant in Poisson's equation in the Newtonian limit
\cite{b11,b12,b13,b14}. In our approach, the effective gravitational
constant is that $\mathcal{G}=\frac{2}{1+\gamma}G_{0}$ for MSTVG
(see Eqs. (37) and (40) in \cite{b8}), where $G_0$ is the {\it bare}
gravitational constant. In other words, the metric component
$g_{00}$ at $\mathcal{O}(1/c^{2})$ is irrelative to the vector
field.

But we do obtain MT's effective gravitational constant from the
motion equation at the Newtonian order. Here we rewrite Eq.(71) in
 DXH as follows
\begin{equation}
   \ddot{x}^i=
  -\frac{\mathcal{G} M}{r^3}x^i
    \left [
    1+\alpha \left ( 1+\frac{r}{\lambda} \right )\exp
   \left ( -\frac {r}{\lambda}\right )\right ] + \mathcal{O}(2).
\end{equation}
If the items within the square brackets are absorbed into the
gravitational constants we nearly get $G_{eff}$ of MT.  This
discussion actually justifies our post-Newtonian approach. On the
other hand we see now that the $G_{eff}$ obtained from the Newtonian
equation of motion is not consistent with the effective
gravitational constant in Poisson equation.

We do not understand why MT said our effective gravitational
constant is $G_N(1+\alpha)$.  In DXH, it is clearly represented in
Eqs.(35) and (40).

 (3) On the value of $\gamma$ for STVG.

 Actually this is the key argument between MT and us. MT wrote
 ``Their argument is based on a Parameterized Post-Newtonian
[4] representation of STVG, which they derive in Appendix B of their
paper by explicitly ignoring the MOG vector field: $\cdots$ ''.

We wish MT had carefully read our paper. The conclusion of
$\gamma=1/3$ is obtained in the context and its detailed derivation
is in Appendix A. In Appendix A we did not ignore the vector field
although the vector field has no contribution to the determination
of $\gamma$ in our approach.

The purpose of Appendix B is to find the values of other PPN
parameters. Because the PPN formalism is restricted within the
gravitational theories which follow EEP \cite{b13} but STVG and
MSTVG violate EEP, the inclusion of the vector field will produce
some super-potentials that are not included in the PPN formalism.
This fact makes us to drop the vector field in Appendix B. The
inclusion of the vector field will not change the fact that
$\gamma=1/3$ for STVG.

(4) On $G\simeq constant$ and $-g_{00}\simeq g_{rr}^{-1}$.

  $G\simeq$const. and $-g_{00}\simeq g^{-1}_{rr}$
 (which denotes $B(r)=A^{-1}(r)$ in \cite{b8}) are used in \cite{b9,b10}.
 In Appendix A of DXH we show that $\gamma=1$ when $G=constant$ but
 otherwise $\gamma=1/3$. Consequently, the sub-family of STVG when
 $G=constant$ can fit the present experiments.

It is well known that $-g_{00}=g^{-1}_{rr}$ means $\gamma=1$
\cite{b13}. For STVG $-g_{00}=g^{-1}_{rr}$ does not hold generally.
In \cite{b9,b10} it is used with $G=constant$ together then it
becomes allowable.

What we considered in DXH is the general STVG theory but not its
special case of $G=constant$. Then $\gamma$ is equal to $1/3$ and
STVG has to be revised to fit the present space experiments.

\newpage 
\bibliography{apssamp}

\end{document}